\documentclass[%
 reprint,
 amsmath,amssymb,
 aps,onecolumn,nofootinbib,
]{revtex4-2}
\usepackage{multirow}
\usepackage{graphicx}
\usepackage{dcolumn}
\usepackage{bm}
\usepackage{fancyhdr}
\usepackage[colorlinks]{hyperref}
\usepackage[nameinlink,noabbrev]{cleveref}
\hypersetup{
    colorlinks=true,
    linkcolor=blue,
    filecolor=magenta,      
    urlcolor=blue,
   citecolor=blue
}
\usepackage{graphicx}
\usepackage{dcolumn}
\usepackage{bm}
\usepackage{scalerel}
\usepackage{tikz}
\usetikzlibrary{svg.path}

\definecolor{orcidlogocol}{HTML}{A6CE39}
\tikzset{
  orcidlogo/.pic={
    \fill[orcidlogocol] svg{M256,128c0,70.7-57.3,128-128,128C57.3,256,0,198.7,0,128C0,57.3,57.3,0,128,0C198.7,0,256,57.3,256,128z};
    \fill[white] svg{M86.3,186.2H70.9V79.1h15.4v48.4V186.2z}
                 svg{M108.9,79.1h41.6c39.6,0,57,28.3,57,53.6c0,27.5-21.5,53.6-56.8,53.6h-41.8V79.1z M124.3,172.4h24.5c34.9,0,42.9-26.5,42.9-39.7c0-21.5-13.7-39.7-43.7-39.7h-23.7V172.4z}
                 svg{M88.7,56.8c0,5.5-4.5,10.1-10.1,10.1c-5.6,0-10.1-4.6-10.1-10.1c0-5.6,4.5-10.1,10.1-10.1C84.2,46.7,88.7,51.3,88.7,56.8z};
  }
}

\newcommand\orcidicon[1]{\href{https://orcid.org/#1}{\mbox{\scalerel*{
\begin{tikzpicture}[yscale=-1,transform shape]
\pic{orcidlogo};
\end{tikzpicture}
}{|}}}}
\usepackage{amsmath}
\usepackage{amssymb, amsfonts}
\usepackage[greek,english]{babel}
\begin{document}

\preprint{APS/123-QED}

\title{Jeans analysis in fractional gravity}


\author{Kamel Ourabah \orcidicon{0000-0003-0515-6728},}\email{kam.ourabah@gmail.com, kourabah@usthb.dz}
\address{Theoretical Physics Laboratory, Faculty of Physics, University of Bab-Ezzouar, USTHB, Boite Postale 32, El Alia, Algiers 16111, Algeria}

\date{\today}

\begin{abstract}
It has recently been demonstrated [A. Giusti, \href{https://doi.org/10.1103/PhysRevD.101.124029}{Phys. Rev. D \textbf{101}, 124029 (2020)}] that characteristic traits of Milgrom's modified Newtonian dynamics (MOND) can be replicated from an entirely distinct framework: a fractional variant of Newtonian mechanics. To further assess its validity, this proposal needs to be tested in relevant astrophysical scenarios. Here, we investigate its implications on Jeans gravitational instability and related phenomena. We examine scenarios involving classical matter confined by gravity and extend our analysis to the quantum domain, through a Schrödinger-Newton approach. We also derive a generalized Lane-Emden equation associated with fractional gravity. Through comparisons between the derived stability criteria and the observed stability of Bok globules, we establish constraints on the theory's parameters to align with observational data.  
\end{abstract}

\maketitle


\section{Introduction}
\label{1}

The success of the standard framework of gravity [read General Relativity (GR) and its weak field limit, namely Newtonian gravity] is widely acknowledged. The validity of Newtonian gravity has been probed at length scales ranging from the millimeter \cite{R00} to the size of planetary orbits \cite{R01}, while GR has been validated by numerous observations \cite{R1,R2} and has even foreseen the existence of new objects, such as black holes \cite{R3}. However, despite such a predictive power, the standard paradigm of gravity is not without its limitations. In fact, to reconcile observations on galactic and cosmological scales, it necessitates the inclusion of two elusive components, namely \textit{dark energy} and \textit{dark matter}. Dark energy is necessary for explaining the accelerated expansion of the universe and is estimated to comprise approximately $70\%$ of its total energy content. Meanwhile, dark matter has been introduced to account for galactic rotation curves, as well as gravitational lensing effects and the large-scale structure of the universe, and it is believed to be around five times more abundant than visible matter. That is to say, in this perspective, ordinary visible matter assumes a very minor role. The failure to detect these dark (yet very abundant) constituents, and to identify their very nature, represents the most significant challenge to the standard framework of gravity \cite{R4,R5,R6,R7}.

These issues have drawn much attention, starting in the 1970s due to the observed rotation curves and the proposed existence of dark matter. Later, in the 1990s, interest intensified with the first evidence of the accelerated expansion of the universe, attributed to dark energy. However, to date, no compelling and consensus solution has emerged. On one front, considerable effort has been directed towards the search for new dark components. On the other front, some authors have shifted their focus to exploring whether novel gravitational physics could account for these discrepancies, without the \textit{ad hoc} inclusion of dark elements \cite{alt0,alt1,alt2,alt3}. After all, the entire basis for the existence of these dark components rests on one premise: the universal applicability of standard gravitational theory across all distance scales. Thus, a clever generalization could eliminate the need for their \textit{ad hoc} introduction.

One of the earliest viable alternatives to dark matter is Milgrom’s modified Newtonian dynamics (MOND) \cite{Milgrom1,Milgrom2,Milgrom3,Milgrom4}. In this theory, Newtonian gravity is modified when the acceleration of a test mass falls below a certain threshold 
$a_0$, which is regarded as a new fundamental constant, empirically determined to be around $10^{-10} m/s^2$. For high accelerations, much above $a_0$, as experienced in the Solar System, MOND reproduces Newtonian gravity, while for extremely low accelerations, as in the outer regions of galaxies, it deviates from Newtonian dynamics in a way that mimics the effect of dark matter. MOND has successfully predicted laws governing galactic dynamics that have been confirmed by observation, but there seem to be specific observations, especially within cosmology, that challenge the theory, at least in its current form (see e.g. \cite{Milgromh} for a review and historical perspective).

Recently, it has been demonstrated \cite{frac}
 that some characteristic effects of MOND can be replicated in a very different setup, namely through a fractional formulation of Newton's theory, using the fractional Poisson equation

\begin{equation}\label{i}
(-\Delta)^s \Phi(\mathbf{r})=-4 \pi G \ell^{2-2 s} \rho(\mathbf{r}),
\end{equation}
where $G$ is the gravitational constant, $\Delta$ represents the standard Laplacian operator, $s$ is a dimensionless parameter such that $1 \leq s \lesssim 3 / 2$, $\Phi(\mathbf{r})$ the (modified) gravitational potential, $\rho (\mathbf{r})$ denotes the distribution of matter density, and $\ell$ is a constant with the dimension of a length. Interestingly, the gravitational potentials satisfying Eq. (\ref{i}) reduce to Newton’s theory for $s=1$, while they reproduce MOND’s large-scale behavior for $s$ approaching the value $3/2$. This trait places fractional gravity as an attractive substitute for dark matter, as it encompasses both Newtonian gravity and MOND's asymptotic behavior. The primary difference between MOND and fractional gravity lies in their fundamental principles\footnote{We note in passing that the potentials satisfying Eq. (\ref{i}) can be derived from an entropic gravity perspective as well \cite{Ourabah2024}, where a fractal (Barrow) entropy is imposed on the cosmological horizon.
}: MOND is an inherently nonlinear theory, characterized by an acceleration scale $a_0$, while fractional gravity is a linear theory, involving a characteristic length scale $\ell$. 

Given the promising prospects of this framework, it becomes essential to explore its applicability in relevant astrophysical scenarios. This paper contributes to this effort by examining how fractional gravity impacts Jeans gravitational instability, and related phenomena. This instability is the mechanism behind the collapse of gravitationally bound systems, such as interstellar gas, ultimately leading to star formation, and it is known to offer interesting possibilities to test different gravitational models and to constraint their parameters \cite{JA1,JA2,JA3,JA4,JA5,JA6}.

The remainder of the paper progresses as follows: In Section \ref{SecII}, we provide a comprehensive overview of the potentials associated with fractional gravity and estimate their impact on the Jeans mass using the virial theorem. In Section \ref{SecIII}, we approach the problem from a hydrodynamic standpoint, and derive generalized Lane-Emden equations associated with fractional gravity. By comparing the derived stability criteria to the observed stability of Bok globules, we establish a condition between the free parameters of the theory, $\ell$ and $s$, to accommodate the data. In Section \ref{SecIV}, we extend our analysis to the quantum realm via a fractional generalization of the Schrödinger-Newton approach. Finally, we summarize our conclusions in Section \ref{SecV}. While the main text adopts a hydrodynamic perspective, an alternative kinetic approach is presented in Appendix A.

\section{Virial equilibrium} \label{SecII}

Using the mathematical definition of the fractional Laplacian operator (see e.g. \cite{Kwa}), the modified gravitational potential satisfying Eq. (\ref{i}) is found as\footnote{Equation (\ref{phis}) can be generalized to $s=3/2$, resulting in the potential $\Phi_{3/2} = \frac{2}{\pi} \frac{G M}{\ell} \log (r / \ell)$, which should be interpreted in the regularized sense (see e.g. \cite{frac,frac2}). However, this transition is far from trivial and will not be considered further here.} \cite{frac}

\begin{equation}\label{phis}
 \Phi_s=   -\frac{\Gamma\left(\frac{3}{2}-s\right) }{4^{s-1} \sqrt{\pi} \Gamma(s)} \left(\frac{\ell}{\mathbf{r}}\right)^{2-2 s} \frac{G m}{\mathbf{r}}, \quad (1 \leq s <3/2). 
\end{equation}

Mathematically speaking, Eq. (\ref{phis}) is nothing but the Green function for the fractional Poisson equation (\ref{i}). Therefore, the potential corresponding to a general density distribution $\rho(\mathbf{r})$ is obtained by a convolution of the latter with Eq. (\ref{phis}) (and  omitting the extra mass $m$).

To gain an initial understanding of how these potential modifications influence the Jeans process, let us derive a stability criterion using a straightforward approach based on the virial theorem. We examine the balance between thermal gas pressure and gravity in an idealized spherical cloud with uniform density $\rho_0$, uniform temperature $T$, and mass $M$. Using Eq. (\ref{phis}), the gravitational binding energy of the spherical cloud reads as 

\begin{equation}\label{W}
|U|=   \int_0^M \Phi_s dm = \frac{3 \Gamma\left(\frac{3}{2}-s\right)}{4^{s-1} \sqrt{\pi} (3+2s) \Gamma(s)}\left(\frac{\ell}{R}\right)^{2-2 s} \frac{G M^2}{R},
\end{equation}
where $R$ is the radius of the spherical cloud. Equation (\ref{W}) reduces to the standard $|U|= 3GM^2/5R$ in the limit $s\to 1$. From another hand, the mean kinetic energy of the gas cloud reads

\begin{equation}\label{K}
K= \frac{3 N k_B T}{2} = \frac{3 M R_g T}{2 \mu},
\end{equation}
where $\mu$ is the mean molecular weight per particle, $k_B$ is the Boltzmann constant, and $R_g \equiv k_B / m_p$. 

An estimate of the Jeans mass, i.e., the minimum mass required for a region of gas to undergo gravitational collapse, can be obtained by applying the virial theorem. The latter states that the kinetic energy of a stable system equals half of its potential energy, i.e., $K = |U| /2$. When this equality holds, the gas cloud is said to be in virial equilibrium. However, when gravity exceeds the outward pressure, the cloud shifts away from virial equilibrium, leading to gravitational collapse. This occurs approximately when $|U| > 2K$. Using Eqs. (\ref{W}) and (\ref{K}), one sees that collapse requires the cloud radius to exceed the threshold value

\begin{equation}\label{R}
R_J^s=\left [  \frac{ \Gamma\left(\frac{3}{2}-s\right) \ell^{2-2s} \mu  G M}{(3+2s) 4^{s-1} \sqrt{\pi} \Gamma(s) R_g T}   \right ]^{1/(3-2s)}.  
\end{equation}

By eliminating the radius in favor of the density $\rho_0$, one obtains the Jeans mass as

\begin{equation}\label{M}
M_J^s = \left [  \left (\frac{3}{4 \pi \rho_0} \right)^{1-2s/3} \frac{(3+2s) 4^{s-1} \sqrt{\pi} \Gamma(s) R_g T}{ \mu G \Gamma\left(\frac{3}{2}-s\right) \ell^{2-2s}}   \right]^{3/2s}.
\end{equation}

Note that the critical mass (\ref{M}) varies with temperature as $\sim T^{3/2s}$ and with density as $\sim \rho_0^{1-3/2s}$. In the limit $s \to 1$, the typical dependencies, namely $\sim T^{3/2}$ and $\sim \rho_0^{-1/2}$, are restored. In this limit, the critical radius (\ref{R}) and mass (\ref{M}) reduce to the (standard) Jeans length and Jeans mass, respectively

\begin{equation}
R_{\mathrm{J}}= \frac{ G M \mu}{5R_{g} T}  \quad \text{and} \quad M_J=\left(\frac{ 5{R_g} {T}}{2 \mu {G}}\right)^{3 / 2}\left(\frac{4}{3} \pi \rho_0 \right)^{-1 / 2}. 
\end{equation}

In order to illustrate the influence of fractional gravity on the Jeans mass in typical astrophysical scenarios, we consider the fundamental components of the interstellar medium (ISM):

\begin{itemize}
\item \textbf{Bok Globules :} 
Small interstellar clouds consisting of extremely cold gas and dust. Their nearly spherical geometry renders them opaque to visible light. They are characterized by a remarkably low temperature ($\sim 10 \text{K}$), a relatively high density ($n \gtrapprox 10^4 \text{cm}^{-3}$), and modest masses (ranging from 1 to 1000 $M_{\odot}$). They are of compact size, typically around $1$pc. In general, they are relatively isolated and often harbor dense cores, making them potential precursors to protostars.

\item \textbf{Giant Molecular Clouds:} Vast conglomerates comprising gas, dust, and complex substructures. Their mass ranges from $10^4 M_{\odot}$ to $10^6 M_{\odot}$, with average density of $ \sim 10^2-10^3 \text{cm}^{-3}$, and a temperature around $10 K$. They extend approximately $10$pc in size. Within GMCs, dense substructures emerge, including Clumps, Dense Cores, and Hot Cores.

\item \textbf{Cold neutral media:}
 Regions composed of neutral hydrogen (HI) and molecules, typically at temperatures ranging from $10K$ to $100K$, with a relatively low number density of about $30 \text{cm}^{-3}$.

  \item \textbf{Warm neutral media:}  Regions consisting of neutral hydrogen (HI) and molecules, typically at temperatures ranging from $10^3K$ to $10^4K$, with a very low number density of around $0.6 \text{cm}^{-3}$.

   \item \textbf{Warm ionized media:} Widespread regions of nearly fully ionized hydrogen. They are characterized by a low density of approximately 0.1 $\text{cm}^{-3}$ and a relatively high temperature of around $8000 \text{K}$. They contribute substantially to the overall interstellar environment.
   
    \item \textbf{HII regions:} Regions of interstellar atomic hydrogen that are ionized, HII regions can take on various shapes due to the irregular distribution of stars and gas within them. Their masses span from $10^2 M_{\odot}$ to $10^4 M_{\odot}$, with particle densities ranging from $0.1 \text{cm}^{-3}$ to $10^4 \text{cm}^{-3}$, and temperatures around $10^4 \text{K}$.

\item \textbf{Hot interclouds:} Regions of ionized gas (HII) with temperatures typically ranging from $10^5$ K to $10^6$ K. These regions represent the lowest density phase in the ISM, with a number density of approximately $0.004 \text{cm}^{-3}$ and a typical size of around $20$pc.

\item \textbf{Fermi Bubbles:} Radiant "8"-shaped structures located on either side of the galactic center of the Milky Way and oriented perpendicular to the galactic disk. Discovered in 2010 by Su \textit{et al.} \cite{Su}, they span a distance of $ \sim 10 \text{kpc}$ and have a temperature ranging from $10^8 K$ to $10^9 K$, with a density of $0.01 \text{cm}^{-3}$.

\end{itemize}

\begin{table}[h]
\centering
\begin{tabular}{ccccccc}
\hline \hline
Phase & $n$ [cm$^{-3}$] & $T$ [K] & $M_J$ [$M_{\odot}$] & $M_J^{s=1.01}$ [$M_{\odot}$] & $M_J^{s=1.1}$ [$M_{\odot}$] & $M_J^{s=1.2}$ [$M_{\odot}$]  \\
\hline 
Bok globules & $10^4$ & 10 & 7.24 & $6.97$ & $5.02$ & $3.55$ \\
Giant molecular clouds & $10^2$ & 10 & 70.7 & $65.2$  & $27.15$  & $11.14$ \\
Cold neutral medium & 30 & 80 & $3 \times 10^3$ &  $2.57 \times 10^3$  & $0.6 \times 10^3$ & $2.09 \times 10^2$ \\
Warm neutral medium & $0.6$ & $8 \times 10^3$ & $1.98 \times 10^7$ & $1.6 \times 10^7$  & $1.6 \times 10^6$ & $1.74 \times 10^5$ \\
Warm ionized medium & $0.1$ & $8 \times 10^3$ & $5.09 \times 10^7$ & $3.8 \times 10^7$  & $2.8 \times 10^6$  & $2.44 \times 10^5$ \\
HII regions & $0.1$ & $10^4$ & $7.07 \times 10^7$ & $5.4 \times 10^7$  & $4.08 \times 10^6$  & $0.35 \times 10^6$ \\
Hot interclouds & $0.004$ & $10^6$ & $3.62 \times 10^{11}$ & $23.7 \times 10^{10}$  & $7.09 \times 10^9$  & $2.44 \times 10^8$ \\
Fermi Bubbles & $0.01$ & $10^8$ & $22.9 \times 10^{13}$ & $14 \times 10^{13}$  & $27 \times 10^{11}$  & $6.27 \times 10^{10}$ \\
\hline \hline
\end{tabular}
\caption{Characteristics of various phases within the ISM alongside the associated Jeans mass under standard and fractional gravity for different values of $s$. The characteristic length $\ell$ is fixed to $10^{15} \text{m}$ for illustrative purposes, while the mean molecular weight $\mu$ is fixed to $2$.}
\label{tab:interstellar}
\end{table}

Table \ref{tab:interstellar} presents the Jeans mass (\ref{M}), for various values of $s$, corresponding to different phases within the ISM. For the sake of illustration, we have chosen a characteristic length of $\ell = 10^{15}$m$\approx 0.032$pc (typically the size of the smallest interstellar clouds). Among the ISM phases, Bok globules are particularly interesting due to their mass being comparable to their Jeans mass. Therefore, they offer a means to distinguish between various gravity theories and to constrain their parameters. In particular, the capacity of fractional gravity to decrease the Jeans mass can elucidate discrepancies observed in the stability of Bok globules \cite{Kandori}, as demonstrated in Section \ref{SecIII}.


\section{Hydrodynamic approach: Classical regime}\label{SecIII}


\subsection{Stability Analysis}

Beyond a simple comparison between kinetic and gravitational binding energies, the Jeans mechanism is best understood through a hydrodynamic perspective, allowing for an examination of perturbation growth rates. For that purpose, we first consider a classical self-gravitating medium, where the self-gravitational potential obeys the fractional Poisson equation (\ref{i}). At the fluid level of description\footnote{Since our primary focus is on determining the stability condition, we confine ourselves to a static universe. Extending this analysis to an expanding universe is straightforward and will be briefly discussed later.}, the dynamics of the medium is governed by the continuity equation

\begin{equation}\label{con}
\frac{\partial \rho}{\partial t}+\nabla \cdot( \rho \mathbf{u})=0,
\end{equation}
where $\rho(\mathbf{r},t)$ is the mass density and $\mathbf{u} (\mathbf{r},t)$ is the velocity field, and the momentum-balance (Euler) equation
\begin{equation}\label{Euler}
 \left[\frac{\partial \mathbf{u}}{\partial t}+(\mathbf{u} \cdot \nabla) \mathbf{u} \right]=-  \nabla \Phi_s- \frac{\nabla p}{\rho},
\end{equation}
where the self-gravitational potential $\Phi_s$ is related to the density through Eq. (\ref{i}). The system of equations (\ref{con}), (\ref{Euler}), and (\ref{i}) is not closed; it requires an equation of state fixing the pressure $p$. Here, we consider the very general scenario of a \textit{barotropic} fluid, where the pressure depends solely on the density. That is,
\begin{equation}
p(\mathbf{r}, t)=p[\rho(\mathbf{r}, t)].
\end{equation}

Equations (\ref{con}), (\ref{Euler}), and (\ref{i}) can be combined in an integro-differential equation as follows

\begin{equation}\label{Eulerid}
 \left[\frac{\partial \mathbf{u}}{\partial t}+(\mathbf{u} \cdot \nabla) \mathbf{u} \right]=-\frac{\nabla }{m}\left [\int \rho(\mathbf{r'}) \Phi_s \left(\left|\mathbf{r}-\mathbf{r}^{\prime}\right|\right) d\mathbf{r}^{\prime}  \right ]- \frac{\nabla p}{\rho},
\end{equation}
where $\Phi_s$ is given by Eq. (\ref{phis}), namely the Green function for Eq. (\ref{i}). Equations (\ref{Eulerid}) and (\ref{con}) can also be expressed in a more compact form as a single equation. That is,
\begin{equation}
 \left[ \frac{\partial ( \rho \mathbf{u})}{\partial t} + \nabla ( \rho \mathbf{u} \otimes \mathbf{u}) \right]= - \frac{\rho \nabla}{m} \left [\int \rho(\mathbf{r'}) \Phi_s \left(\left|\mathbf{r}-\mathbf{r}^{\prime}\right|\right) d\mathbf{r}^{\prime}  \right ]- {\nabla p}.
\end{equation}

We confine our analysis to the linear regime, focusing on small perturbations within a stationary, infinite, homogeneous, and isotropic equilibrium medium, and write

\begin{equation}
\rho (\mathbf{r},t)= \rho_0+ \delta \rho (\mathbf{r},t) \quad \text{and} \quad  \mathbf{u}(\mathbf{r},t)= \mathbf{u}_0+ \delta \mathbf{u}(\mathbf{r},t),
\end{equation}
where $\delta \rho$ and $\delta \mathbf{u}$ are supposed small. Upon linearizing the fluid equations (\ref{con}) and (\ref{Eulerid}), we obtain\footnote{Note that here we are assuming that the self-potential arises solely from the perturbation $\delta \rho$ and not from the background density 
$\rho_0$. This practice is commonly known as the "Jeans swindle". In fact, the Jeans approach to self-gravitating systems is known to face a mathematical inconsistency from the outset. This inconsistency arises from the fact that, according to the Poisson equation, a constant background potential implies a vanishing density $\rho_0 =0$. Jeans \cite{Jeans} resolved this inconsistency by assuming that the Poisson equation is sourced \textit{only} by the perturbation and \textit{not} by the background density $\rho_0$. Although this may seem \textit{ad hoc}, it has been demonstrated to be a mathematically rigorous procedure (see e.g., \cite{J1,J2}). In our setup, this can be avoided by adding a background potential $V_0$ to the momentum-balance equation (\ref{Euler}), such that
\begin{equation*}
V_0 =- \rho_0  \left [\int  \Phi_s \left(\left|\mathbf{r}-\mathbf{r}^{\prime}\right|\right) d\mathbf{r}^{\prime}  \right ],
\end{equation*}
which would play a role similar to that of the electric potential generated by the ionic background in a plasma when analyzing electron waves (plasmons). Alternatively, one may circumvent the Jeans swindle by considering the expansion of the universe \cite{J3}, or by examining the dynamical stability of an inhomogeneous distribution of matter within a finite domain \cite{J4}.}
\begin{equation}\label{H1}
\begin{aligned}
\frac{\partial \delta \rho (\mathbf{r},t)}{\partial t}&+ \rho_0 \nabla \cdot \delta \mathbf{u} (\mathbf{r},t) =0,\\
\frac{\partial \delta \mathbf{u} (\mathbf{r},t)}{\partial t}&=  -\frac{\nabla}{m} \left [\int \delta \rho(\mathbf{r'},t) \Phi_s \left(\left|\mathbf{r}-\mathbf{r}^{\prime}\right|\right) d\mathbf{r}^{\prime}  \right ]-\frac{c_s^2 \nabla \delta \rho (\mathbf{r},t)}{\rho_0},
\end{aligned}
\end{equation}
where 
\begin{equation}\label{cs}
c_{s}^{2}=  \left(\frac{d p}{d \rho} \right)_{\rho=\rho_0}
\end{equation}
is the squared speed of sound in the medium. Performing a Fourier transform of Eq. (\ref{H1}) and combining the resulting equations (see e.g., \cite{Ourabah2023}), one arrives at the following dispersion relation
\begin{equation}\label{gdr}
\omega^{2}=\left(\rho_{0} / m\right) \widetilde{\Phi_s}(k) k^{2} + c_s^2 k^2,
\end{equation}
where 
\begin{equation}\label{Vk}
\widetilde{\Phi_s}(k)=\int \Phi_s(r) \exp [-i \mathbf{k} \cdot \mathbf{r}] d \mathbf{r}
\end{equation}
is the Fourier transform of the self-potential $\Phi_s$ [i.e., Eq. (\ref{phis})]. Using the identity

\begin{equation}
\frac{1}{r^\alpha} = \frac{2\pi^\alpha}{\Gamma(\alpha/2)} \int_{0}^{\infty} \lambda^{\alpha-1} e^{-\pi \lambda^2 r^2} d\lambda,
\end{equation}
which is easy to Fourier transform, by interchanging the order of integration, namely

\begin{equation}
\int_{\Bbb R^n} e^{-\pi \lambda^2 |x|^2} e^{- i k \cdot x} dx = \lambda^{-n} e^{-\pi |k|^2/ 4 \pi \lambda^2},
\end{equation}
one obtains

\begin{equation}\label{phisk}
\widetilde{\Phi_s}(k)= -  \frac{4 \pi G m \ell^{2-2s}}{k^{2s}}.
\end{equation}
In $3d$, the latter is valid as long as $s<3/2$. Replacing Eq. (\ref{phisk}) into Eq. (\ref{gdr}), one obtains the dispersion relation for a gravitational medium in fractional gravity. That is,

\begin{equation}\label{dr}
w^2 = - \Omega_J^2 (\ell k)^{2-2s} + c_s^2 k^2,
\end{equation}
where $\Omega_J= \sqrt{4 \pi G \rho_0}$ is the so-called Jeans frequency.

As known, the instability manifests itself through the imaginary part of the frequency. Thus, let us express the frequency as $\omega \equiv \omega_r + i \omega_i$. Note that, because $\omega^2$ is real, $\omega$ is either real or purely imaginary. The condition $\omega=0$ determines a critical wave-number, which delineates between the two scenarios. According to the dispersion relation (\ref{dr}), such a critical wave-number is given by

\begin{equation}\label{kstar}
 k^*= \frac{1}{\ell} \left( \frac{\ell \Omega_J}{c_s} \right)^{1/s},
\end{equation}
which reduces to the (standard) Jeans wave-number, $k_J= \Omega_J / c_s$, in the limit $s=1$. For $k>k^*$, one has $\omega=\omega_r=\pm \sqrt{\omega^2}$, so that the perturbation behaves as $\sim \exp(-i \omega_r t)$ with a real frequency

\begin{equation}
\omega_r = \sqrt{c_s^2k^2- \Omega_J^2 (\ell k)^{2-2s}},
\end{equation}
without attenuation. This scenario represents gravity-modified sound waves. In that case the system is stable. Conversely, for $k<k^*$, one has $\omega= \omega_i= \pm i \sqrt{- \omega^2}$ so that the perturbation grows (without oscillation) like $\sim \exp(\omega_i t)$, with a growth rate given by

\begin{equation}
\omega_i=\sqrt{\Omega_J^2 (\ell k)^{2-2s}-c_s^2 k^2}. 
\end{equation}
In that case, the system is unstable; it experiences gravitational collapse. Introducing the following dimensionless variables,
\begin{equation}\label{dim}
 \overline{\gamma} := \frac{\omega_i}{\Omega_J}, \quad \overline{k}:= \frac{k}{k_J}, \quad \text{and} \quad \alpha := \ell k_J, 
\end{equation}
the growth rate can be expressed in a dimensionless form as

\begin{equation}\label{GRa}
\overline{\gamma} = \left [(\alpha \overline{k})^{2-2s} - \overline{k}^2  \right]^{1/2}.
\end{equation}

\begin{figure*}
\centering
\includegraphics[width=0.5\linewidth]{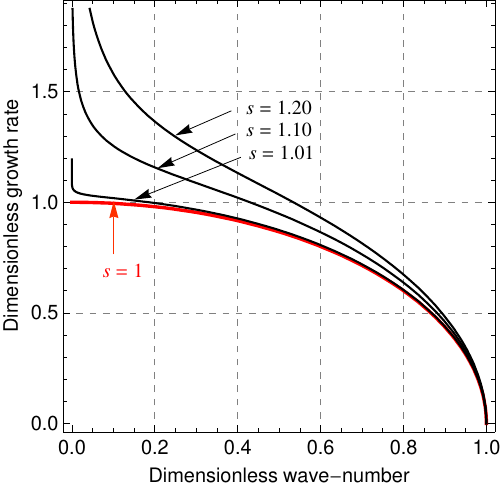}
\caption{The dimensionless growth rate $\overline{\gamma}:= \omega_i / \Omega_J$ [viz. Eq. (\ref{GRa})] as a function of the dimensionless wave number $\overline{k}:= k/ k_J$, for $\ell k_J =1$.} \label{FigGRa}
\end{figure*}

The dimensionless growth rate is depicted in Fig. \ref{FigGRa}, for different values of $s$. One may observe that, as in Newtonian gravity, the growth rate reaches its maximum for $k=0$ (infinite wavelengths). However, unlike Newtonian gravity where the maximum growth rate is given by $\sqrt{4 \pi G \rho_0}$, in fractional gravity, the growth rate increases as $s$ departs from unity and diverges in the infinite wave-lengths limit ($k \to 0$).

From the critical wave-number (\ref{kstar}), one may define a critical mass, defined as the mass contained in a sphere of radius $2 \pi / k^*$ with a density $\rho_0$. It reads as 
\begin{equation}\label{Mc}
M^*= \frac{32 \pi^4 \rho_0 \ell^3}{3} \left ( \frac{c_s}{\Omega_J \ell} \right)^{3/s},
\end{equation}
which is nearly identical to Eq. (\ref{M}), derived from the virial theorem, but gives a better estimate for the Jeans process in a barotropic fluid. In the limit $s=1$, it reduces to the standard Jeans mass of a barotropic fluid, namely
\begin{equation}\label{MJ}
M_J=  \frac{32 \pi^4 \rho_0}{3} \left ( \frac{c_s}{\Omega_J } \right)^{3}.
\end{equation}
Comparing the critical mass (\ref{Mc}) with the standard Jeans mass (\ref{MJ}), one may observe that, for $\ell < c_s/\Omega_J$, one has $M^*<M_J$, leading to gravitational collapse for smaller masses compared to Newtonian gravity. Conversely, for $\ell > c_s/\Omega_J$, the situation is the other way around.

Finally, it is worth noting that while we are considering the case of a static universe, which is sufficient for determining the stability criterion, extending the study to an expanding universe is straightforward (see e.g. \cite{Bonnor}). In this case, the density contrast $\delta (t):= \delta \rho (t)/ \rho_0$ is determined by

\begin{equation}\label{Bc}
\ddot{\delta}+2 \frac{\dot{a}}{a} \dot{\delta}+\left[\frac{c_s^2 k^2}{a^2}- \Omega_J^2 \left(\frac{\ell k}{a} \right)^{2-2s}\right] \delta=0,
\end{equation}
where $a(t)$ is the scale factor. Equation (\ref{Bc}) can be regarded as an extension of Bonnor equation \cite{Bonnor} to fractional gravity. In a static universe ($a=1$), the dispersion relation (\ref{dr}) is recovered upon writing $\delta \propto \exp(-i \omega t)$.

\subsection{Data Analysis}

One way to assess the physical viability of the theory is to test with data of regions in the universe that can experience star formation. In this respect, Bok globules offer an excellent testing environment because their mass closely aligns with their Jeans mass. As a result, even a slight deviation in the stability condition can yield a different prediction regarding their stability.

In Table \ref{tab:globules}, we reproduce the data of 11 Bok globules sourced from \cite{Kandori} (see also \cite{Vainio}). One may observe that, for 7 of these globules, the observed stability contradicts the expected stability condition: they are observed to be unstable (undergoing star formation) despite having a mass smaller than their Jeans mass. This discrepancy has been approached from various perspectives, with some attributing it to the influence of dark matter \cite{DM1,DM2}. Fractional gravity's ability to mimic the effects of dark matter presents an interesting avenue for investigating its potential role in resolving this discrepancy. 
As initially noted in Ref. \cite{JA3}, reducing the critical mass by a factor of $2/5$, that is
\begin{equation}
M^*= \left(\frac{2}{5} \right) M_J,
\end{equation}
ensures the correct stability for the Bok globules presented in Table \ref{tab:globules}. By employing this phenomenological factor as a sufficient condition to match with the observed stability of Bok globules, we can derive a saturation bound for the parameters $\ell$ and $s$, which reads as

\begin{equation}\label{sat}
\frac{\ell \Omega_J}{c_s} \leq \left ( \frac{2}{5} \right )^{\frac{s}{3(s-1)}}.  
\end{equation}

\begin{figure}[h]
\centering
\includegraphics[width=0.5\linewidth]{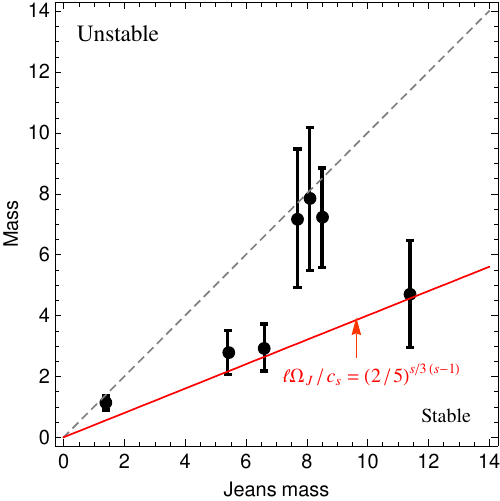}
\caption{Mass and Jeans mass of 7 Bok globules listed in Table \ref{tab:globules}, where the predicted stability contradicts observations. The dashed gray line distinguishes between the stable and collapsing regions according to the standard Jeans criterion ($s=1$). The solid red line represents the threshold of $2/5$, providing a sufficient condition to account for the data.} \label{FigBG}
\end{figure}

That is, for $\ell$ and $s$ satisfying this condition, fractional gravity successfully accounts for the discrepancy observed in the stability of Bok globules. For illustration, Figure \ref{FigBG} shows the data for 7 Bok globules from Table \ref{tab:globules}, where the observed stability contradicts the predicted stability, together with the saturation bound (\ref{sat}) for reference.

\begin{table}[h!]
\scriptsize
\begin{tabular}{cccccc} \hline \hline
Bok Globule &$T\text{[K]}$ & $n\text{[cm$^{-3}$]}$ & $M [M_{\odot}]$  & $M_J [M_{\odot}]$ & Stability \\ \hline
CB 87	& 11.4 &	$(1.7\pm 0.2)\times 10^4$	& $2.73\pm 0.24$ & 9.6 & stable \\ 
CB 110 & 21.8 & $(1.5\pm 0.6)\times 10^5$ & $7.21\pm 1.64$ & 8.5  & unstable \\ \
CB 131 & 25.1 & $(2.5\pm 1.3)\times 10^5$ & $7.83\pm 2.35$ & 8.1  & unstable \\ 
CB 134 & 13.2 & $(7.5\pm 3.3)\times 10^5$ & $1.91\pm 0.52$ & 1.8 & unstable \\ 
CB 161 & 12.5 & $(7.0\pm 1.6)\times 10^4$ & $2.79\pm 0.72$ & 5.4   & unstable \\ 
CB 184 & 15.5 & $(3.0\pm 0.4)\times 10^4$ & $4.70\pm 1.76$ & 11.4 & unstable \\ 
CB 188 & 19.0 & $(1.2\pm 0.2)\times 10^5$ & $7.19\pm 2.28$ & 7.7  & unstable \\ 
FeSt 1-457 & 10.9 & $(6.5\pm 1.7)\times 10^5$ & $1.12\pm 0.23$ & 1.4   & unstable \\ 
Lynds 495 & 12.6 & $(4.8\pm 1.4)\times 10^4$ & $2.95\pm 0.77$ & 6.6  & unstable \\ 
Lynds 498 & 11.0 & $(4.3\pm 0.5)\times 10^4$ & $1.42\pm 0.16$ & 5.7  & stable \\ 
Coalsack & 15 & $(5.4\pm 1.4)\times 10^4$ & $4.50$ & 8.1  & stable \\ \hline \hline
\end{tabular}
\caption{Temperature, particle number density, mass, Jeans mass, and observed stability of multiple Bok globules, documented in \cite{Kandori}.}
\label{tab:globules}
\end{table}

\subsection{Lane-Emden Equation}

From the hydrodynamic set of equations, one may derive a generalized Lane-Emden equation, describing a self-gravitating, spherically symmetric, polytropic fluid. To do this, one may observe that the stationary solutions of Eq. (\ref{Euler}) satisfy the condition

\begin{equation}\label{eq0}
\nabla p = - \rho \nabla \Phi_s,
\end{equation}
which can be viewed as a condition of hydrostatic equilibrium. For a spherically symmetric distribution, one has
\begin{equation}\label{eq}
\frac{d p}{d r}=- \frac{(2s-3) \Gamma(3/2-s) G M(r) \rho(r) }{4^{s-1} \sqrt{\pi} \Gamma(s)} \left(\frac{\ell}{r}\right)^{2-2s}.
\end{equation}
For a polytropic fluid, the pressure reads as\footnote{The equation of state (\ref{poly}) enables the modeling of various scenarios of astrophysical interest. For example, as $n \to \infty$ ($\gamma=1$), it converges to the isothermal equation of state. For a monoatomic gas or a completely degenerate gas of electrons obeying the Fermi-Dirac statistics at absolute zero temperature, one has \cite{Chandrasekhar1942} $\gamma=5/3$ ($n=3/2$). Additionally, an approximate polytropic equation of state with an index of $\gamma \approx 3.25$ ($n \approx 0.44$) is applicable in the radiative region of a star \cite{Chavanis2010}.}
\begin{equation}\label{poly}
p=K \rho^\gamma=K \rho^{(n+1) / n},
\end{equation}
where $K$ is a constant and $\gamma \equiv 1+ 1/n$, with $n$ being the \textit{polytropic index}. The polytropic equation of state (\ref{poly}) describes an adiabatic transformation where the specific entropy remains constant. In this context, the adiabatic index $\gamma$ denotes the ratio of specific heats at constant pressure and constant volume, i.e. $\gamma = c_p/c_v$. Combining Eqs. (\ref{eq}) and (\ref{poly}) with the continuity condition,
\begin{equation}
\frac{d M}{d r}=4 \pi r^2 \rho(r),
\end{equation}
one has
\begin{equation}\label{LED}
\frac{1}{r^2} \frac{d}{d r}\left(\frac{r^{4-2s} K}{\rho} \gamma \rho^{\gamma-1} \frac{d \rho}{d r}\right)=- \frac{(2s-3) \Gamma(3/2-s) \ell^{2-2s} 4 \pi G \rho}{4^{s-1} \sqrt{\pi} \Gamma(s)}. 
\end{equation}

We rescale the radial variable and the density as
\begin{equation}
 r \equiv \alpha \xi \quad \text{and} \quad  \rho(r)=\rho_c \theta^n(\xi),
\end{equation}
where $\xi$ and $\theta$ are dimensionless variables, and $\rho_c$ is the central density. Upon defining $\alpha$ as
\begin{equation}
\alpha \equiv \left ( \frac{4^{s-1} \sqrt{\pi} \Gamma(s) K (n+1) \rho_c^{\frac{1}{n}-1}  }{4 \pi G (2s-3) \Gamma(3/2-s) \ell^{2-2s}}  \right )^{1/(4-s)},
\end{equation}
Equation (\ref{LED}) gives the following Lane-Emden equation

\begin{equation}\label{LE}
\frac{1}{\xi^2} \frac{d}{d \xi}\left(\xi^{4-2s} \frac{d \theta}{d \xi}\right)=-\theta^n,
\end{equation}
which reduces to the standard Lane-Emden equation for $s=1$, as required. For an arbitrary polytropic index $n$, Equation (\ref{LE}) does not admit a closed-form solution, but it is interesting to note the following homology transformation: if $\theta(\xi)$ is a solution to Eq. (\ref{LE}), then so is $ A^{2s/(n-1)}\theta (A \xi)$.

Note that at the center ($\xi=0$), the density corresponds to $\rho_c$, thus $\theta(0)=1$. Moreover, as $dp/dr$ approaches $0$ as $r \to 0$, we have $d \theta/dr|_{\xi=0} = 0$. These constitute the boundary conditions for the solution. The outer boundary, or the surface, is the first location $\xi_1$ where $\rho=0$ (or equivalently $\theta=0$). While the formal solution may have additional zeros at larger values of $\xi >\xi_1$, they are not relevant for stellar models.

\begin{figure*}
\centering
\begin{minipage}[t]{0.3\linewidth}
\includegraphics[width=1\linewidth]{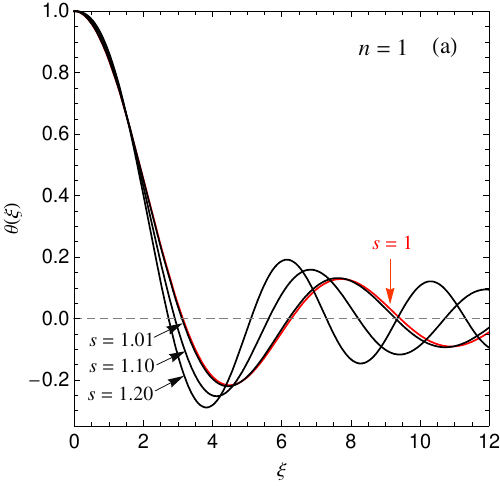}
\end{minipage}
\begin{minipage}[t]{0.3\linewidth}
\includegraphics[width=1\linewidth]{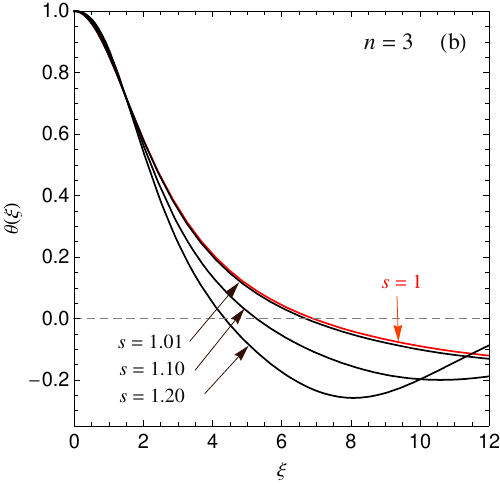}
\end{minipage}
\begin{minipage}[t]{0.3\linewidth}
\includegraphics[width=1\linewidth]{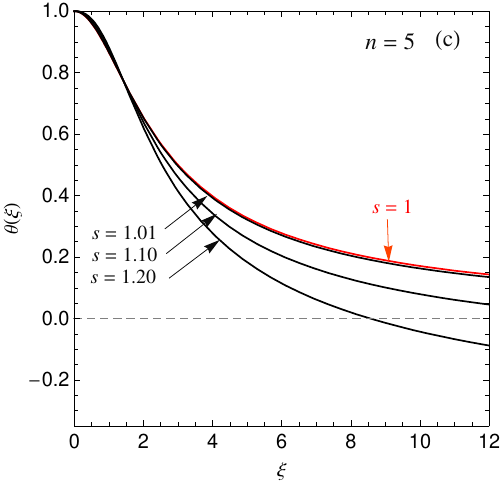}
\end{minipage}
\caption{Numerical solutions $\theta (\xi)$ to the Lane-Emden equation (\ref{LE}) for different values of $s$ and different polytropic indices, $n=1$, $n=3$, and $n=5$.} \label{Fig2}
\end{figure*}

Figure \ref{Fig2} shows numerical solutions of the Lane-Emden equation (\ref{LE}) for various values of $s$ and different polytropic indices $n=1$, $3$, and $5$. Note that $n=3$ corresponds to the standard Eddington model \cite{Eddington}, representing a relativistic degenerate gas, particularly relevant for neutron stars and highly massive white dwarfs. The figure shows a trend: as the parameter $s$ moves away from unity,  it results in smaller radii for stellar structures, as compared to those predicted by Newtonian gravity.

\section{Hydrodynamic approach: Quantum Regime}\label{SecIV}

We now turn our attention to addressing the same problem within the quantum realm. To model such a scenario, one typically begins with the Schrödinger-Newton (SN) equation, also known as the Schrödinger-Poisson equation. This equation couples the Schrödinger equation, which describes the evolution of the wave function $\psi$, with the Poisson equation governing the variation of the gravitational potential. In the context of fractional gravity, the model can be expressed as 

\begin{equation}\label{SN}
\begin{aligned}
&i \hbar \frac{\partial \psi}{\partial t}=-\frac{\hbar^2}{2m} \Delta \psi+ m \Phi_s \psi, \\
&-(\Delta)^s \Phi_s=- 4 \pi G \ell^{2-2s}|\psi|^2,
\end{aligned}
\end{equation}

or, equivalently, as an integro-differential equation as follows
\begin{equation}\label{idd}
\mathrm{i} h \frac{\partial \psi}{\partial t}=\left[-\frac{\hbar^{2}}{2 m} \Delta-m  \int {\Phi_s (\left|\mathbf{r}-\mathbf{r}^{\prime}\right|) \left|\psi\left(\mathbf{r}^{\prime}, t\right)\right|^{2}} \mathrm{dr}^{\prime}\right] \psi,
\end{equation}
where the self-potential $\Phi_s$ is given by Eq. (\ref{phis}). Initially, the SN equation was introduced independently by Diósi \cite{SN1}, Penrose \cite{SN2}, and others \cite{TT1,TT2} as a mechanism to explain the collapse of the wave function due to gravitational effects. In the astrophysics and cosmology literature, it is the standard framework to describe dense matter confined by gravitational interactions, as for example in boson stars \cite{BS1,BS2,BS3} and scalar field dark matter \cite{SFDM1,SFDM2,SFDM3,SFDM4,Ourabah2020bis,Our2020}. Given the implications of the original SN equation in addressing conceptual problems in quantum mechanics\footnote{For instance, numerical simulations of the (original) SN equation \cite{SNp1,SNp2,SNp3} have shown that gravitational effects start affecting the Schrödinger dynamics beyond a critical mass $m_c \approx {\bar{h}^2}/{R^{2/3}G^{-1}}
$, where $R$ is the size of the object. Simulations of gravitational effects in the SN model, using optical set-ups playing the role of 'gravity analogs', have also been considered \cite{O1,O2}.}, the fractional generalization (\ref{SN}) could have profound consequences at the fundamental level. Yet, we defer these inquiries here and focus instead on exploring the consequences of this equation on the Jeans mechanism.

\subsection{Homology Analysis}

Before exploring the consequences of the model (\ref{SN}) on the Jeans mechanism, let us take a moment to briefly discuss the homology or scale transformation in play in this context. To do so, let us rescale the four variables
\begin{equation}
\psi \rightarrow A \psi, \quad \Phi_s \rightarrow A^a \Phi_s, \quad r \rightarrow A^b r, \quad \text { and } \quad t \rightarrow A^c t,
\end{equation}
and determine the values of the exponents that yield the original equations (\ref{SN}). This results in the following equations for the exponents



\begin{equation}
1-c=1-2 b=1+a  \quad \text{and} \quad a-2 b s=2,
\end{equation}
and, consequently, in the following scaling forthe variables

\begin{equation}
\psi \rightarrow A \psi, \quad  \Phi_s \rightarrow A^{2/1+s} \Phi_s, \quad r \rightarrow A^{-1 / 1+s} r,\quad \text { and } \quad t \rightarrow A^{-2 / 1+s} t. 
\end{equation}

In purely gravitational problems, it is common practice to normalize the wave function such that $|\psi(\textbf{r},t)|^2=\rho (\textbf{r},t)$. In this case, the conserved total mass (assuming spherical symetry) is expressed as
\begin{equation}
M=\int_0^{\infty} 4 \pi r^2|\psi|^2 d r,
\end{equation}
which scales as 
\begin{equation}
    M \to A^{2-{3}/{(1+s)}} M.
\end{equation}
For $s=1$, one recovers the standard scaling of the SN equation $M \to A ^{1/2}M$, while in the limit $s \to 3/2$, it approaches the scaling $M \to A^{4/5} M$.

\subsection{Stability Analysis}

To study the implications of Eq. (\ref{SN}) on the Jeans instability of quantum gravitational media, it is appropriate to transform the SN equation into a set of hydrodynamic equations. For that purpose, we employ the so-called Madelung transformation \cite{Madelung}, and write the wave function $\psi$ in polar form

\begin{equation}\label{polar}
\psi(\mathbf{r}, t)=  A(\mathbf{r}, t) e^{i S(\mathbf{r}, t) / \hbar},
\end{equation}
where $A(\mathbf{r}, t)$ and $S(\mathbf{r}, t)=(\hbar / 2 i) \ln \left(\psi / \psi^{*}\right)$ are real functions, representing the amplitude and the phase of the wave function. The mass density and the velocity field are defined in terms of $A$ and $S$ as

\begin{equation}
\rho(\mathbf{r}, t)=  |\psi|^{2}=   A(\mathbf{r}, t)^2 \quad \text { and } \quad \mathbf{u}=\frac{\nabla S}{m }=\frac{i \hbar}{2 m } \frac{\psi \nabla \psi^{*}-\psi^{*} \nabla \psi}{|\psi|^{2}}.
\end{equation}
Note that, as defined, the velocity field is irrotational, i.e., $\nabla \times \mathbf{u}=\mathbf{0}$. By substituting the wave function (\ref{polar}) into Eq. (\ref{idd}) and separating the real and imaginary parts, we obtain the continuity equation from the imaginary part
\begin{equation}\label{QH1}
\frac{\partial \rho}{\partial t}+\nabla \cdot(\rho \mathbf{u})=0,
\end{equation}
while the real part leads to
\begin{equation}\label{8}
\frac{\partial S}{\partial t}+\frac{1}{2 m}(\nabla S)^{2}+ \frac{1}{m}\int A^2(\mathbf{r'}) \Phi_s \left(\left|\mathbf{r}-\mathbf{r}^{\prime}\right|\right) d\mathbf{r}^{\prime}+Q=0,
\end{equation}
where
\begin{equation}
Q \equiv -\frac{\hbar^{2}}{2 m} \frac{\Delta \sqrt{\rho}}{\sqrt{\rho}}=-\frac{\hbar^{2}}{4 m}\left[\frac{\Delta \rho}{\rho}-\frac{1}{2} \frac{(\nabla \rho)^{2}}{\rho^{2}}\right]
\end{equation}
is the so-called quantum potential or the Bohm potential. Taking the gradient of Eq. (\ref{8}), and considering that $\nabla \times \mathbf{u}=\mathbf{0}$, we arrive at the momentum-balance equation
\begin{equation}\label{QH2}
 \left[\frac{\partial \mathbf{u}}{\partial t}+(\mathbf{u} \cdot \nabla) \mathbf{u} \right]=-\frac{\nabla}{m} \left [\int \rho(\mathbf{r'}) \Phi_s \left(\left|\mathbf{r}-\mathbf{r}^{\prime}\right|\right) d\mathbf{r}^{\prime}  \right ]- \frac{\nabla Q}{m}.
\end{equation}
Equations (\ref{QH1}) and (\ref{QH2}) form the quantum hydrodynamic set of equations. Proceeding similarly to Section \ref{SecIII} for the classical regime, we derive the following dispersion relation\footnote{Here again, in an expanding universe, the density contrast $\delta (t) := \delta \rho (t) / \rho_0$ is determined by
\begin{equation*}
\ddot{\delta}+2 \frac{\dot{a}}{a} \dot{\delta}+\left[\frac{\hbar^2 k^4}{4 m^2 a^4}- \Omega_J^2 \left(\frac{\ell k}{a} \right)^{2-2s}\right] \delta=0.
\end{equation*}
The dispersion relation (\ref{dr2}) is recovered for $a=1$ with $\delta \propto \exp(-i \omega t)$.}

\begin{equation}\label{dr2}
w^2 = - \Omega_J^2 (\ell k)^{2-2s} + \frac{\hbar^2}{4 m^2} k^4,
\end{equation}
which is similar to the classical dispersion relation (\ref{dr}), except that the dispersion term scales as $\sim k^4$. Equation (\ref{dr2}) describes the interplay between the attractive gravitational attraction and quantum pressure forces that tend to stabilize the Jeans instability for large values of $k$, or short
wavelengths. As in the classical regime, the critical wave-number delimiting between stable and unstable regions corresponds to $\omega^2=0$. In this case, it reads as

\begin{equation}
k^*= \frac{1}{\ell} \left ( \frac{2 m \ell^{2} \Omega_J}{\hbar} \right )^{\frac{1}{1+s}},
\end{equation}
which reduces to the standard result
\begin{equation}\label{kjq}
k_J= \left ( \frac{2m \Omega_J}{\hbar} \right)^{1/2},
\end{equation}
for $s=1$. The associated critical mass, i.e., the mass contained in a sphere of radius $2 \pi / k^*$, reads

\begin{equation}\label{Mq}
M^*= \frac{32 \pi^4 \rho_0 \ell^3}{3} \left ( \frac{\hbar}{2m \ell^{2} \Omega_J}\right )^{\frac{3}{1+s}}.
\end{equation}

It represent the minimum mass of a fluctuation that can undergo gravitational collapse. Therefore, it provides an order of magnitude estimation for the minimum mass of quantum structures confined by gravitational fields. Note that in the context of dark matter, Eq. (\ref{Mq}) with $s=1$ corresponds to the Jeans mass of bosonic dark matter without collisions\footnote{This model is known by various names, including wave dark matter, fuzzy dark matter, BECDM, $\psi$DM, among others. See \cite{SFDM3} for a historical overview and \cite{Rr} for a comprehensive review.}. For non-interacting bosons, in order to reproduce the scales of dark matter halos, the dark matter particle is estimated to be around $m \sim 10^{-22}$eV. Such an ultra-light particle does not correspond to any known particles and is in tension with observations of the Lyman-$\alpha$ forest \cite{alpha}. In this context, fractional gravity might offer a solution by accommodating dark matter particles with a higher mass. Of course, the primary aim of alternative gravitational theories is to eliminate the \textit{ad hoc} introduction of dark matter, they may however not entirely obviate its necessity but can address some of its challenges. For instance, it seems that \cite{beta} MOND fails to fully eliminate the need for dark matter in all astrophysical systems; residual mass discrepancies persist, particularly evident in galaxy clusters even under MOND analysis.

Using the dimensionless variables (\ref{dim}), with the standard Jeans wave-number $k_J$ given by (\ref{kjq}), the dimensionless growth rate can be expressed as

\begin{equation}\label{GRb}
\overline{\gamma} = \left [ (\alpha \overline{k})^{2-2s} - \overline{k}^4 \right]^{1/2}.
\end{equation}

\begin{figure*}
\centering
\includegraphics[width=0.5\linewidth]{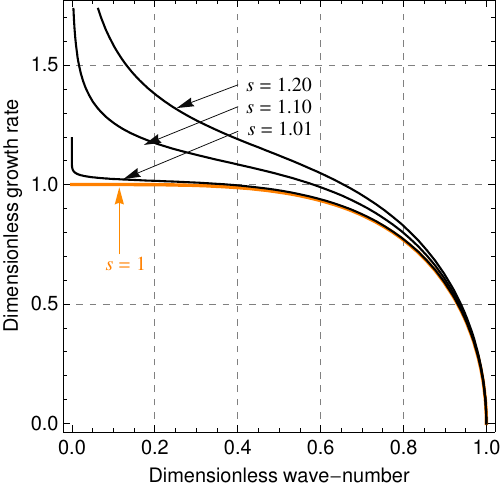}
\caption{The dimensionless growth rate $\overline{\gamma}:= \omega_i / \Omega_J$ [viz. Eq. (\ref{GRb})] as a function of the dimensionless wave-number $\overline{k}:= k/ k_J$, for $\ell k_J =1$.} \label{FigGRb}
\end{figure*}

The latter is depicted in Fig. \ref{FigGRb}. As in the classical regime, the growth rate increases as $s$ deviates from unity and diverges as wave-lengths approach infinity ($k \to 0$), in contrast to Newtonian gravity where the maximum growth rate is given by $\sqrt{4 \pi G \rho_0}$.

\section{Conclusion}\label{SecV}

This study investigates some phenomenological implications of fractional gravity, a concept recently proposed by Giusti \cite{frac}. This framework offers a fresh perspective by reproducing some characteristic features of Milgrom's modified Newtonian dynamics (MOND), within the context of a fractional variant of Newtonian mechanics. While MOND is a fundamentally non-linear theory characterized by an acceleration scale $a_0$, fractional gravity is a linear theory, involving a characteristic length parameter $\ell$. Our analysis focuses on the impact of fractional gravity on the Jeans gravitational instability for classical and quantum media. Furthermore, we derive generalized Lane-Emden equations associated to fractional gravity, shedding light on its broader implications. Through comparisons  between the derived stability criteria and the observed stability of
Bok globules, we established constraints on the theory’s parameters to account for the observational data.

This analysis opens up interesting possibilities regarding 'gravity analogs', namely systems capable of emulating gravitational phenomena in laboratory settings. Particularly in the quantum realm, one can take advantage of the formal analogies between self-gravitating systems and non-gravitational media, which exhibit similar elementary excitations. Examples include plasmons in quantum plasmas, Bogoliubov excitations in Bose-Einstein condensates, and hybrid phonon modes in magneto-optical traps \cite{Ourabah2023,Tito2,moi}. Notably, in the latter medium, dispersion properties identical to those of quantum matter confined by Newtonian gravity can be replicated \cite{T1,T3}. Emulating the dispersion relations of fractional gravity (\ref{dr2}) may be realized through the manipulation of radiation and laser absorption cross-sections.

\appendix
\section*{Appendix A: Kinetic Approach}\label{Ap}

In the main text, the problem is addressed in the language of hydrodynamics. Here, we offer an alternative kinetic treatment of the problem. While we focus on a quantum kinetic (wave kinetic) approach, based on the Wigner approach, the results of a classical kinetic approach (based on the Vlasov equation) can be straightforwardly recovered by taking the formal limit $\hbar \to 0$. We start by defining the Wigner function
\renewcommand{\thesection}{A.\arabic{section}}
\setcounter{equation}{0}

\begin{equation}\label{Wf}
W(\mathbf{r}, \mathbf{q}, t)=\int \psi^*(\mathbf{r}-\mathbf{s} / 2, t) \psi(\mathbf{r}+\mathbf{s} / 2, t) \exp (i \mathbf{q} \cdot \mathbf{s}) \mathrm{d} s,
\end{equation}
which is simply the Fourier transform of the auto-correlation function corresponding to the wave-function $\psi$. Here, it is normalized such that
\begin{equation}
\int W(\mathbf{r}, \mathbf{q}, t)  \frac{d \mathbf{q}}{( 2 \pi)^3}=\left|\psi(\mathbf{r}, t)\right|^{2} = \rho(\mathbf{r},t),
\end{equation}
where $\rho(\mathbf{r},t)$ denotes the mass density.
Applying the well-known Wigner–Moyal procedure \cite{Wigner,Moyal}, the generalized SN equation (\ref{idd}) can be rewritten as (see for instance \cite{Tito,Ourabahlivre} for detailed calculations)

\begin{equation}\label{Wigner}
\mathrm{i} \hbar\left(\frac{\partial}{\partial t}+\mathbf{v}_q \cdot \nabla\right) W=  \int \widetilde{\Phi_s}(\boldsymbol{\kappa}) \widetilde{ \rho}(\boldsymbol{\kappa}, t) \Delta W \mathrm{e}^{i \boldsymbol{\kappa} \mathrm{r}} \frac{\mathrm{d} \boldsymbol{\kappa}}{(2 \pi)^3},
\end{equation}
where $\mathbf{v_q} \equiv \hbar \mathbf{q} /m$ is the particle velocity and $\hbar \mathbf{q}$ is its momentum and we have defined $\Delta W=\left[W^{-}-W^{+}\right]$ with $W^{ \pm}=W(\mathrm{r}, \mathrm{q} \pm \kappa / 2, t)$. Above,

\begin{equation}
\widetilde{\rho} (\kappa, t)=\int \rho (\mathbf{r}, t) \exp (-\mathrm{i} \kappa \cdot \mathbf{r}) \mathrm{d} \mathbf{r}
\end{equation}
is the Fourier transform of $ \rho (\mathbf{r},t)$ and $\widetilde{\Phi_s}$ is the Fourier transform of the self-potential $\Phi_s$ [viz. Eq. (\ref{phisk})]. Equation (\ref{Wigner}) is the quantum analog of the Vlasov (collisionless Boltzmann) kinetic equation, to which it reduces in the formal limit $\hbar \to 0$. It can be approached similarly to the hydrodynamic set of equations; by considering small density perturbations and expressing $W = W_0 + \delta W$, where $W_0$ denotes an equilibrium state and $\delta W$ represents the perturbed quantity, which is assumed to evolve spatially and temporally as $\sim \exp(i \mathbf{k} \cdot \mathbf{r} - i \omega t)$. Upon linearizing the Wigner equation (\ref{Wigner}), one obtains

\begin{equation}
\delta {W}= {\widetilde{\Phi_s}(\mathbf{k})} \frac{\Delta W_0}{\hbar\left(\omega-\mathbf{k} \cdot \mathbf{v}_q\right)} \Tilde{\rho}(\mathbf{k}).
\end{equation}
Integrating over $\mathbf{q}$,one obtains the quantum kinetic dispersion relation as

\begin{equation}\label{ird}
1- {\widetilde{\Phi_s}(\mathbf{k})} \int \frac{\Delta W_0}{\hbar\left(\omega-\mathbf{k}-\mathbf{v}_q\right)} \frac{\mathrm{dq}}{(2 \pi)^3}=0,
\end{equation}
which can be rewritten as follows
\begin{equation}
1-\frac{\widetilde{\Phi_s}(\mathbf{k})}{\hbar }  \int W_0(\mathbf{q})\left[\frac{1}{\left(\omega_{+}-\mathbf{k} \cdot \mathbf{v}_q\right)}-\frac{1}{\left(\omega_{-}-\mathbf{k} \cdot \mathbf{v}_q\right)}\right] \frac{\mathrm{d} \mathbf{q}}{(2 \pi)^3}=0,
\end{equation}
with $\omega_{ \pm} \equiv \omega \pm \hbar k^2 / 2 m$. Redefining the equilibrium distribution as the projected (marginal) distribution along the parallel axis, i.e.,

\begin{equation}
W_0(q) \to \int W_0\left(q, \mathbf{q}_{\perp}\right) \frac{\mathrm{d} \mathbf{q}_{\perp}}{(2 \pi)^2},
\end{equation}
Equation (\ref{ird}) becomes
\begin{equation}\label{aaa}
1-\frac{ \widetilde{\Phi_s}(\mathbf{k}) k^2}{m \omega^2}  \int \frac{W_0(u) \mathrm{d} u}{(1-k u / \omega)^2-\hbar^2 k^4 / 4 m^2 \omega^2}=0,
\end{equation}
where $u \equiv q/m$ is the velocity. Considering the long wave-length limit $u << \omega /k$, and assuming an even distribution $W_0$, which is characteristic of equilibrium and nearly equilibrium situations, Equation (\ref{aaa}) yields
\begin{equation}\label{rdqt}
\omega^2 \simeq-\Omega_J^2 (\ell k)^{2-2s} +3 k^2\left\langle u^2\right\rangle+\frac{\hbar^2 k^4}{4 m^2},
\end{equation}
which remains valid as long thermal and quantum corrections are small. Equation (\ref{rdqt}) interpolates between the pure quantum dispersion relation (\ref{dr2}) and the thermal dispersion relation (\ref{dr}), with $c_s^2 \equiv 3 \langle u^2 \rangle$, discussed in the main text. The pure quantum dispersion relation (\ref{dr2}) emerges when kinetic effects are disregarded, meaning the distribution shrinks to a Dirac delta distribution, i.e., $W_0(q)= \rho_0 \delta (q)$. Conversely, the thermal dispersion relation (\ref{dr}) is recovered in the formal limit $\hbar \to 0$, or more precisely for long wave-lengths (small $k$).

\section*{Data availability statement}
No new datasets were generated for this study. Data from \cite{Kandori} were used for Table \ref{tab:globules} and Figure \ref{FigBG}.

\end{document}